\documentclass[12pt,a4paper]{article}

\usepackage{amsmath,amssymb,epsfig,graphics}
\allowdisplaybreaks

\def\d{\text{d}}
\def\e{\text{e}}

\def\Sm{\Sigma_-}
\def\Sc{\Sigma_\times}
\def\Nm{N_-}
\def\Nc{N_\times}
\def\Sp{\Sigma_+}

\begin{document}
\begin{center}
{\Large{\bf Numerical confirmations of joint spike transitions in $G_2$ cosmologies}}

\

{\bf W C Lim}

\

Department of Mathematics, University of Waikato, Private Bag 3105, Hamilton 3240, New Zealand

\

wclim@waikato.ac.nz
\end{center}

\begin{abstract}
We produce numerical evidence that the joint spike transitions between Kasner eras of $G_2$ 
cosmologies are described by the non-orthogonally transitive $G_2$ spike solution.
A new matching procedure is developed for this purpose.
\end{abstract}


\section{Introduction}

Belinskii, Khalatnikov and Lifschitz~\cite{art:LK63,art:BKL1970,art:BKL1982} describe the dynamics of a spacetime (with fluids with soft equation of state)
as it evolves toward a generic spacelike singularity as an infinite sequence of transitions between anisotropic vacuum Kasner saddle states. 
The transitions are described by vacuum Bianchi type II solutions. 
The time spent near a Kasner state is called a Kasner epoch. Each Kasner state is characterised by its BKL parameter $u\geq 1$,
which decreases by 1 for each Bianchi type II transition. 
A Kasner era consists of a sequence of Kasner epochs with decreasing BKL parameter $u$.
If $u<2$ before the transition, then $u$ is mapped to $\frac{1}{u-1}$ after the transition, starting a new Kasner era.

Due to spatial inhomogeneities, neighbouring worldlines 
experience slightly different, but increasingly diverging, sequences.
The dynamics are asymptotically local, as if each typical worldline follows the dynamics of a spatially homogeneous model.
Exception to this locality was found when spikes were discovered numerically by Berger and Moncrief~\cite{art:BergerMoncrief1993}
in the context of vacuum models with two commuting Killing vector fields that act orthogonally transitively (OT $G_2$ models).
Spike forms along worldlines where the crucial variable responsible for the Bianchi type II transition ($N_-$ in our formulation below) is zero, 
as it changes sign due to spatial inhomogeneities~\cite{art:ColeyLim2014}.
When a spike forms, its width can become temporarily narrower than the particle horizon, and its dynamics is described by a spatially inhomogeneous solution.
The exact OT $G_2$ spike solution was found in 2008~\cite{art:Lim2008}, and is shown to match non-moving spikes in numerical simulations in 2009~\cite{art:Limetal2009}, 
which introduces a zooming technique as a cost-effective way to maintain adequate numerical resolution. 
The exact solution describes a transition between Kasner states that are connected by two consecutive Bianchi type II transitions.
OT $G_2$ models allow only a single Kasner era plus the first Kasner epoch of the next era, so a spike forming near the end of the era ends up as a permanent spike.
Relaxing the orthogonal transitivity condition allows multiple Kasner eras and a non-terminating sequence of transitions.
The permanent spike is replaced by the so-called joint spike transition~\cite{art:HeinzleUgglaLim2012}, 
which is a more elaborate spike transition that straddles two Kasner eras.
The results of ~\cite{art:Limetal2009} is therefore incomplete until a new exact solution describing the joint spike transition is found and numerically matched.

The exact non-OT $G_2$ spike solution that seems to describe such a transition was found in 2015~\cite{art:Lim2015}.
However, the exact non-OT $G_2$ spike solution uses a different time parameterisation,
and it was not known at the time how to recover this time variable in numerical simulations.
In~\cite{art:ColeyLim2016}, where the exact solution is generalised to the stiff fluid case,
the fluid-comoving or volume gauge seemed to be a more natural gauge choice than the area time gauge.
The numerical simulation in~\cite{art:ColeyLim2016}
uses the fluid-comoving gauge in conjunction with a dynamic version of the zooming technique (developed in~\cite{art:LimRegisClarkson2013}).
Doing so comes at a cost of fine-tuning the excision boundary.
The precision of fine-tuning is determined by the size of the spatial domain at the end of simulation relative to that at the start.
Fine-tuning requires one to run numerical simulations multiple times.
Therefore, while the goal to extend the result of the paper~\cite{art:Limetal2009} can be achieved, it would come at a great cost, which is very inefficient and unsatisfactory.
Since 2017, much effort has been spent on improving the fine-tuning through better understanding of the transition times of exact solution.
A method of analysing the transition times was developed in the doctoral thesis of Moughal~\cite{thesis:Moughal2021,art:MoughalLim2021},
and it was then applied to the non-OT $G_2$ spike solution in~\cite{art:LimMoughal2022}.
Despite the better understanding, improvement in the fine-tuning is very little.
Attempts to modify the zooming technique to avoid the fine-tuning problem have been unsuccessful.

The breakthrough came in 2021 when it was realised that we can actually keep using the area time gauge and \emph{evolve the time variable of the exact solution} to accommodate numerical matching,
thus avoiding fine-tuning.
This paper will show how this is done, develop a new matching procedure, and show that the exact solution indeed matches the joint spike transitions.
The reader should read the paper~\cite{art:Limetal2009} and make frequent comparisons as the two papers are similar in the approach.

\section{$G_2$ spacetimes}

We use the orthonormal frame approach~\cite{book:WainwrightEllis1997,art:Ugglaetal2003} to formulate the Einstein field equations,
adopting the Iwasawa spatial frame~\cite{art:HeinzleUgglaRohr2009} and $\beta$-normalised variables~\cite{art:vEUW2002} for numerical evolution (although we will plot some Hubble-normalised variables).
We represent the metric components of non-OT $G_2$ models the same way as in~\cite{art:Lim2015}, where indices
$0$, $1$, $2$, $3$ correspond to coordinates $t$, $x$, $y$, $z$, and the metric components
are given in terms of $N$, $b^1$, $b^2$, $b^3$, $n_1$, $n_2$, $n_3$ as follows:
\begin{align}
	g_{00} &= - N^2,
\\
	g_{11} &= \e^{-2b^1},\quad g_{12} = \e^{-2b^1} n_1,\quad g_{13} = \e^{-2b^1} n_2,
\\
	g_{22} &= \e^{-2b^2} + \e^{-2b^1} n_1{}^2,\quad g_{23} = \e^{-2b^1} n_1 n_2 + \e^{-2b^2} n_3,
\\
	g_{33} &= \e^{-2b^3} + \e^{-2b^1} n_2{}^2 + \e^{-2b^2} n_3{}^2.
\end{align}
The metric components depend on $t$ and $z$ only. Note the change in alignment and notation from $(\tau,x)$ in~\cite{art:Limetal2009}
to $(t,z)$ here.

The expansion shear components $\Sigma_{\alpha\beta}$ and remaining nonzero spatial curvature components $N_{\alpha\beta}$ here are decomposed as follows:
\begin{align}
        \Sigma_{\alpha\beta}
        &= \left(\begin{matrix}
		\Sp + \sqrt{3} \Sm	& \sqrt{3} \Sc 		& \sqrt{3} \Sigma_2 \\
		\sqrt{3} \Sc		& \Sp - \sqrt{3} \Sm 	& \sqrt{3} \Sigma_1 \\
		\sqrt{3} \Sigma_2	& \sqrt{3} \Sigma_1	& -2\Sp
	\end{matrix}\right),
\\
	N_{\alpha\beta}
        &= \left(\begin{matrix}
                2\sqrt{3} \Nm	& \sqrt{3} \Nc	& 0 \\
                 \sqrt{3} \Nc	& 0		& 0 \\
                0		& 0		& 0
        \end{matrix}\right).
\end{align}
We use the frame rotation freedom to set $\Sigma_2=0$.

The $\beta$-normalised variables in terms of the metric components are:
\begin{gather}
	\beta = -\frac12 N^{-1} \partial_t (b^1+b^2),\quad	\mathcal{N} = N \beta,\quad	E_3{}^3 = \frac{\e^{b^3}}{\beta},
\\
	\Sp = \frac13 ( 1 + \mathcal{N}^{-1} \partial_t b^3 ),\quad	\Sigma_{1} = \frac{\e^{b^2-b^3}}{2\sqrt{3}} \mathcal{N}^{-1} \partial_t n_3,
\\
	\Sm = -\frac{1}{2\sqrt{3}} \mathcal{N}^{-1} \partial_t (b^1-b^2),\quad	\Nc = \frac{1}{2\sqrt{3}} E_3{}^3 \partial_z (b^1-b^2),
\\
	\Sc = \frac{\e^{b^1-b^2}}{2\sqrt{3}} \mathcal{N}^{-1} \partial_t n_1,\quad	\Nm = \frac{\e^{b^2-b^1}}{2\sqrt{3}} E_3{}^3 \partial_z n_1,
\end{gather}
where $\partial_t$ and $\partial_z$ denote partial differentiation with respect to $t$ and $z$ respectively.
$t$ tends to infinity toward the singularity.
The Hubble expansion scalar $H$ is related to $\beta$ through $H = \beta(1-\Sp)$,
so Hubble-normalised variables (denoted with a superscript $H$) are related to $\beta$-normalised variables through
\begin{equation}
	\Sp^H = \frac{\Sp}{1-\Sp},\quad \Sm^H = \frac{\Sm}{1-\Sp},
\end{equation}
and so on.

We use the same temporal gauge and time parameterisation for numerical evolution as in~\cite{art:Limetal2009}, namely the area time gauge~\cite{art:vEUW2002} and the time parameterisation such that
\begin{equation}
	\mathcal{N} = -\frac12.
\end{equation}
To maintain adequate numerical resolution as spikes become narrow when they form,
we use the same zooming technique as in~\cite{art:Limetal2009}, introducing zooming coordinates $(T,Z)$ to zoom in on a specified worldline $z=z_\text{zoom}$:
\begin{equation}
	T = t,\quad Z = \frac{2}{(E_3^3)_0}\e^t (z-z_\text{zoom}).
\end{equation}
Compare with Equation (20) of~\cite{art:Limetal2009}.
In this paper we allow $(E_3{}^3)_0$ (the initial value of $E_3{}^3$) to take value other than $2$.
We fix the zoom rate to the natural zoom rate ($A=1$ in Equation (20) of~\cite{art:Limetal2009}) here because we can now better estimate the growth of the lower bound on the right excision boundary, 
which we will elaborate on below. 
The differential operators in the new coordinates are
\begin{equation}
	\partial_t = \partial_T + Z \partial_Z,\quad	\partial_z = \frac{2}{(E_3^3)_0} \e^T \partial_Z.
\end{equation}
The evolution equations in the zooming coordinates are
\begin{align}
\label{dt_E33}
	\partial_T E_3{}^3 &= -Z \partial_Z E_3{}^3 - (1-\tfrac32\Sigma_1^2) E_3{}^3
\\
\label{dt_Sm}
	\partial_T \Sm &= -Z \partial_Z \Sm + \e^T \frac{E_3{}^3}{(E_3{}^3)_0} \partial_Z \Nc + \frac32\Sigma_1^2\Sm - \sqrt{3}(\Sc^2-\Nm^2) + \frac{\sqrt{3}}{2}\Sigma_1^2
\\
	\partial_T \Nc &= -Z \partial_Z \Nc + \e^T \frac{E_3{}^3}{(E_3{}^3)_0} \partial_Z \Sm - (1-\tfrac32\Sigma_1^2)\Nc
\\
	\partial_T \Sc &= -Z \partial_Z \Sc - \e^T \frac{E_3{}^3}{(E_3{}^3)_0} \partial_Z \Nm + (\tfrac32\Sigma_1^2 + \sqrt{3}\Sm) \Sc + \sqrt{3}\Nc\Nm
\\
\label{dt_Nm}
	\partial_T \Nm &= -Z \partial_Z \Nm - \e^T \frac{E_3{}^3}{(E_3{}^3)_0} \partial_Z \Sc - (1-\tfrac32\Sigma_1^2-\sqrt{3}\Sm) \Nm - \sqrt{3}\Nc\Sc
\\
\label{dt_Sigma_1}
	\partial_T \Sigma_1 &= -Z \partial_Z \Sigma_1 + \tfrac12(3\Sigma_1^2+3\Sp-\sqrt{3}\Sm)\Sigma_1
\intertext{where}
\label{Sp}
	\Sp &= \tfrac12(1-\Sm^2-\Sc^2-\Sigma_1^2-\Nm^2-\Nc^2).
\end{align}
For numerical accuracy, we choose to evolve the logarithm of $E_3{}^3$ and $\Sigma_1$.
There is one constraint equation:
\begin{equation}
\label{constraint_eq}
	\e^T \frac{E_3{}^3}{(E_3{}^3)_0} \partial_Z \Sigma_1 = (3\Nm\Sc-3\Nc\Sm-\sqrt{3}\Nc) \Sigma_1.
\end{equation}
Equations~(\ref{dt_E33})--(\ref{constraint_eq}) are essentially the same as Equations~(22)--(28) of~\cite{art:Limetal2009}.

The characteristic velocity for the evolution equations (\ref{dt_E33}) and (\ref{dt_Sigma_1}) is $Z$.
To find the characteristic velocities for the subsystem (\ref{dt_Sm})--(\ref{dt_Nm}), write them in the form
\begin{align}
        \partial_T (\Sm + \Nc) &= -(Z - \e^T \frac{E_3{}^3}{(E_3{}^3)_0}) \partial_Z (\Sm + \Nc) + \cdots,
\\
        \partial_T (\Sm - \Nc) &= -(Z + \e^T \frac{E_3{}^3}{(E_3{}^3)_0}) \partial_Z (\Sm - \Nc) + \cdots,
\\
        \partial_T (\Sc - \Nm) &= -(Z - \e^T \frac{E_3{}^3}{(E_3{}^3)_0}) \partial_Z (\Sc - \Nm) + \cdots,
\\
        \partial_T (\Sc + \Nm) &= -(Z + \e^T \frac{E_3{}^3}{(E_3{}^3)_0}) \partial_Z (\Sc + \Nm) + \cdots.
\end{align}
Then we see that their characteristic velocities are $Z \pm \e^T \frac{E_3{}^3}{(E_3{}^3)_0}$.
So the maximum and minimum characteristic velocities of the system (\ref{dt_E33})--(\ref{dt_Sigma_1}) are
\begin{equation}
\label{v_max_min}
	v_\text{max} = Z + \e^T \frac{E_3{}^3}{(E_3{}^3)_0},\quad
	v_\text{min} = Z - \e^T \frac{E_3{}^3}{(E_3{}^3)_0}.
\end{equation}
To avoid specifying boundary conditions, we want to ensure that all characteristic velocities are outgoing at the excision boundary.
So we want to place the excision boundary far away enough from $Z=0$, so that
$v_\text{min} \geq 0$ at the right excision boundary $Z=Z_r$ and $v_\text{max} \leq 0$ at the left excision boundary $Z=Z_l$
throughout the duration of numerical evolution.
How does $E_3{}^3$ behave?
The joint spike transition undergoes one $\Sigma_1$ frame transition, during which $\Sigma_1$ is of order 1. $\Sigma_1$ is otherwise negligible.
From numerical observations, we see that
\begin{equation}
	E_3{}^3 \approx (E_3{}^3)_0 \e^{-T}
\end{equation}
before the $\Sigma_1$ transition, and
\begin{equation}
        E_3{}^3 \approx \frac{4}{(w-1)^2} (E_3{}^3)_0 \e^{-T}
\end{equation}
after the $\Sigma_1$ transition (for the case $0 < w < 1$, as we will make this choice in~(\ref{case_choice}) below), where $w$ is a parameter of the exact spike solution below.
This means $v_\text{min}$ drops from $Z-1$ to $Z-\frac{4}{(w-1)^2}$. That is, the lower bound for the right-hand excision boundary,
obtained by solving $v_\text{min} =0$ for $Z$ in~(\ref{v_max_min}):
\begin{equation}
\label{Z_right_bound}
	Z_\text{right bound} = \e^T \frac{E_3{}^3}{(E_3{}^3)_0},
\end{equation}
grows from $Z_\text{right bound}=1$ to $Z_\text{right bound}=\frac{4}{(w-1)^2}$.
Therefore the right excision boundary $Z=Z_r$ must satisfy $Z_r \geq \frac{4}{(w-1)^2}$.
Similarly, the left excision boundary $Z=Z_l$ must satisfy $Z_l \leq -\frac{4}{(w-1)^2}$.

In the matching procedure below, the value of $w$ can be obtained before the $\Sigma_1$ transition occurs, thus the excision boundary can be easily adjusted after a first run.
If $w$ is too close to $1$, then $Z_\text{right bound}$ is too large and the numerical simulation becomes too expensive to run, and one really needs a more sophisticated zooming technique.
If the numerical run covers a second $\Sigma_1$ transition into the third Kasner era, then $Z_\text{right bound}$ will increase a second time, 
and again a more sophisticated zooming technique should be used to reduce wastage of numerical resources.
We shall leave that to future research.

The Courant-Friedrichs-Lax condition for numerical stability requires that the numerical timestep size $\Delta T$ satisfies
\begin{equation}
	\Delta T < \frac{\Delta Z}{v_\text{max}},
\end{equation}
evaluated at the excision boundary, where $\Delta Z$ is the numerical grid size. We use
\begin{equation}
        \Delta T = 0.9 \frac{\Delta Z}{v_\text{max}}.
\end{equation}
We use the same numerical method as in~\cite{art:Limetal2009}, namely the classical fourth-order Runge-Kutta method, with fourth-order accurate spatial derivatives.
We use double precision for our numerical runs, and use quad precision to check that our double-precision runs are accurate.

\section{The exact non-OT $G_2$ spike solution}

The exact non-OT $G_2$ spike solution in~\cite{art:Lim2015} with $K=0$ and $\omega_0=0$ is given by:
\begin{align}
\label{nonOT_spike}
        N &= -\e^{-\frac14(w^2+3)\tau} \sqrt{\omega^2+\lambda^2}
\\
        \e^{-2b^1} &= \frac{\lambda}{\omega^2+\lambda^2}
\\
        \e^{-2b^2} &= \frac{\mathcal{A}^2}{\lambda} (\omega^2+\lambda^2)
\\
\label{nonOT_spike_b3}
        \e^{-2b^3} &= \e^{-\frac12(w^2+3)\tau} \mathcal{A}^{-2} (\omega^2+\lambda^2)
\\
        n_1 &= -2w(w-1) n_{30} z^2
%\notag\\
%        &\quad
                + \frac{\omega^2}{\lambda}(n_{30} \e^{-(w+1)\tau} + n_{10} n_{20} \e^{-\frac12(w^2-1)\tau})
\notag\\
        &\quad  -\Bigg[ n_{30} w \e^{-2\tau} + n_{10} n_{20} \frac{(w+3)}{(w-1)} \e^{-\frac12(w-1)^2\tau}
\notag\\
        &\quad\qquad
                + n_{20} n_{30} (n_{10} n_{30} - n_{20}) \frac{(w-3)}{(w+1)} \e^{-\frac12(w+1)^2\tau} \Bigg]
\\
        n_2 &= n_{20} \Bigg[ \frac{\omega^2}{\lambda} \e^{-\frac12(w^2-1)\tau} - \frac{(w+3)}{(w-1)} \e^{-\frac12(w-1)^2\tau}
%\notag\\
%        &\quad\qquad
                - n_{30}^2 \frac{(w-3)}{(w+1)} \e^{-\frac12(w+1)^2\tau} \Bigg]
\label{nonOT_spike_n2}
\\
        n_3 &= \mathcal{A}^{-2} \left[ n_{10} \e^{-\frac12(w-1)^2\tau} + n_{30} (n_{10} n_{30} - n_{20}) \e^{-\frac12(w+1)^2\tau} \right],
\label{nonOT_spike_n3}
\intertext{where}
        \mathcal{A}^2 &= \e^{-2\tau} + n_{10}^2 \e^{-\frac12(w-1)^2\tau} + (n_{10} n_{30} - n_{20})^2 \e^{-\frac12(w+1)^2\tau}
\\
        \lambda &= \e^{(w-1)\tau} + n_{20}^2 \e^{-\frac12(w^2-1)\tau} + n_{30}^2 \e^{-(w+1)\tau}
\\
        \omega &= 2w n_{30} z.
\end{align}
To keep $\Sigma_2=0$, the parameters must satisfy
\begin{equation}
\label{n20_restriction}
	n_{20} = \frac{4w}{(w-1)(w+3)} n_{10} n_{30}.
\end{equation}
Note that the exact solution uses a different time parameterisation $\tau$, with
\begin{equation}
\label{volume_gauge}
	N = - \e^{-b^1-b^2-b^3},
\end{equation}
and not $N=-\frac{1}{2\beta}$ as used in the numerical evolution.
As mentioned in the introduction, using the gauge condition~(\ref{volume_gauge}) in numerical simulation would come at a cost of fine-tuning.
To avoid this, we shall keep using the area time gauge in numerical simulation and evolve $\tau$ along the spike worldline through~(\ref{dtau_dt}) below to accommodate numerical matching.

\section{Matching with exact solution}

A new matching procedure is needed.
We often rescale and shift the coordinates variables to simplify the exact solution, but the numerical solution does not necessarily appear in the simplified form.
To accommodate numerical matching, we relax the parameterisation of the coordinates for the exact solution, to linear order:
\begin{equation}
	\tau \rightarrow k_0(\tau-\tau_0),\quad
	x \rightarrow k_1 x,\quad
	y \rightarrow k_2 y,\quad
	z \rightarrow k_3 z,
\end{equation}
where $k_0$, $k_1$, $k_2$, $k_3$ are positive constants, and $\tau_0$ is a constant.
This has the following effect on the metric components:
\begin{equation}
	N \rightarrow k_0 N,\quad
	b^\alpha \rightarrow b^\alpha - \ln k_\alpha,\quad
	n_1 \rightarrow \frac{k_2}{k_1} n_1,\quad
	n_2 \rightarrow \frac{k_3}{k_1} n_2,\quad
	n_3 \rightarrow \frac{k_3}{k_2} n_3.
\end{equation}
But we shall maintain the time parameterisation~(\ref{volume_gauge}), leading to the relation
\begin{equation}
\label{k_relation}
	k_0 = k_1 k_2 k_3.
\end{equation}
The spike solution is generated by the Geroch transformation, 
which leaves arbitrary additive functions of $y$ and $z$ in $\beta_2$ and $\beta_3$, and hence in $n_1$ and $n_2$ (see~\cite[Equations (27)--(29)]{art:Lim2015}).
But because zooming reduces spatial dependence to essentially linear order, and the fact that we will require the metric components to be even functions of $Z$ in the section below,
adding a constant in $n_1$ and $n_2$ will suffice in most cases.
To accommodate numerical matching, we must also evolve the time variable $\tau$ of the exact solution along the spike worldline $Z=0$, with evolution equation
\begin{equation}
\label{dtau_dt}
	\frac{\d \tau}{\d T} = \frac{N_\text{numerical}}{N_\text{exact}} = \frac{\e^{b^1+b^2+b^3}}{2\beta} = \frac{\e^{b^1+b^2}}{2} E_3{}^3.
\end{equation}
We do not need to recover $\tau$ at places other than the spike worldline, because the matching procedure below performs the matching \emph{only along the spike worldline}.

For completeness we shall evolve all the spatial metric components, which will allow us to match every metric component.
Their evolution equations in zooming coordinates are%
\footnote{We take this opportunity to correct the errors in~\cite[Equations (C1a)--(C1b)]{art:HeinzleUgglaLim2012}). See~\cite[Equation (A.11b)]{art:HeinzleUgglaRohr2009} for the correct equations.}
\begin{align}
	\partial_T b^1 &= -Z \partial_Z b^1 + \tfrac12(1+\sqrt{3}\Sm)
\\
	\partial_T b^2 &= -Z \partial_Z b^2 + \tfrac12(1-\sqrt{3}\Sm)
\\
	\partial_T b^3 &= -Z \partial_Z b^3 + \tfrac12(1-3\Sp)
\\
	\partial_T n_1 &= -Z \partial_Z n_1 - \sqrt{3} \Sc \e^{b^1-b^2}
\\
        \partial_T n_2 &= -Z \partial_Z n_2 - \sqrt{3} \Sc \e^{b^1-b^2} n_3
\\
\label{dt_n3}
	\partial_T n_3 &= -Z \partial_Z n_3 - \sqrt{3} \Sigma_1 \e^{b^2-b^3}.
\end{align}
$\beta$ and $\Sp$ have their own evolution equations, but we do not need them since we can compute $\beta$ and $\Sp$ algebraically using $\beta = \e^{b^3}/ E_3{}^3$ and~(\ref{Sp}).

From~\cite{art:Lim2015}, we see that the exact spike solution is multiply-represented, where the same state-space orbit yields multiple values of $w$.
Here we shall give the new matching procedure using a small positive $w$, with
\begin{equation}
\label{case_choice}
		0 < w < 1. 
\end{equation}
Similar matching procedures can be developed for the other cases.
We shall focus on the joint spike transition only (described by the first alternative in~\cite{art:Lim2015}), ignoring the uninteresting second alternative in~\cite{art:Lim2015},
which describes what is essentially an OT $G_2$ spike transition preceded and succeeded by $\Sigma_1$ frame transitions. 
The new matching procedure to determine the values of parameters $w$, $n_{10}$, $n_{20}$, $n_{30}$, $\tau_0$, $k_0$, $k_1$, $k_2$, $k_3$ is as follows.

Firstly, we determine the value of the parameter $w$.
We plot the combination
\begin{equation}
\label{sm_sc_combination}
	s_1(T) = (\Sm + \tfrac{2}{\sqrt{3}})^2 + \Sc^2
\end{equation}
along $Z=0$. For a joint spike transition, this combination behaves like a sigmoid curve, transitioning from the value $\tfrac13 w^2$ to the value $\frac{(w+3)^2}{3(w-1)^2}$.
We thus obtain an approximate value for $w$ from $s_1$.

Next, the combination
\begin{equation}
\label{temp}
	s_2(\tau) = (-6 b^1 -2 b^2 -2 b^3)|_{Z=0} = -\frac12(w^2+3) k_0 (\tau -\tau_0) + \ln(k_0^2 k_1^4)
\end{equation}
allows us to obtain $k_0$ through its slope against $\tau$.

Thirdly, we want to match $e^{2b^1}|_{Z=0}$ with
\begin{align}
	&\frac{1}{k_1^2} \left[ \e^{(w-1)k_0(\tau-\tau_0)} + n_{20}^2 \e^{-\frac12(w^2-1)k_0(\tau-\tau_0)} + n_{30}^2 \e^{-(w+1)k_0(\tau-\tau_0)} \right]
\\
	&= C_1 \e^{(w-1)k_0 \tau} + C_2 \e^{-\frac12(w^2-1)k_0\tau} + C_3 \e^{-(w+1)k_0 \tau},
\end{align}
where the coefficients $C_1$ , $C_2$ and $C_3$ are given by
\begin{equation}
\label{C1C2C3}
	C_1 = \frac{1}{k_1^2} \e^{(w-1)k_0(-\tau_0)},\
	C_2 = \frac{1}{k_1^2} n_{20}^2 \e^{-\frac12(w^2-1)k_0(-\tau_0)},\
	C_3 = \frac{1}{k_1^2} n_{30}^2 \e^{-(w+1)k_0(-\tau_0)}.
\end{equation}
That is, we want to minimise the relative difference
\begin{equation}
	M(C_1,C_2,C_3) = \sum \frac{(e^{2b^1}|_{Z=0} - C_1 \e^{(w-1)k_0 \tau} - C_2 \e^{-\frac12(w^2-1)k_0\tau} - C_3 \e^{-(w+1)k_0 \tau})^2}{(e^{2b^1}|_{Z=0})^2},
\end{equation}
where the sum is done over selected numerical data points.
$M$ is a sum of squares and is quadratic in $C_1$, $C_2$ and $C_3$ with a single critical point, which is a local minimum point.
Its global minimum point is located at the critical point, following the method of least squares.
To find the critical point, we solve the system
\begin{equation}
	\frac{\partial M}{\partial C_1} = 0,\quad
	\frac{\partial M}{\partial C_2} = 0,\quad
	\frac{\partial M}{\partial C_3} = 0.
\end{equation}
This yields the (unique) values for $C_1$, $C_2$ and $C_3$.
$\ln C_1^2$ plus a particular value for $s_2(\tau)$ eliminates $k_1$, and can be solved to give $\tau_0$:
\begin{equation}
\label{tau0}
\tau_0 = \frac{\ln C_1^2 + (-6 b^1 -2 b^2 -2 b^3)|_{Z=0} - \ln k_0^2 + \frac12(w^2+3)k_0 \tau}{ [\tfrac12(w^2+3) -2(w-1)]k_0}.
\end{equation}
Then $k_1$, $|n_{20}|$ and $|n_{30}|$ are obtained from $C_1$, $C_2$ and $C_3$ through~(\ref{C1C2C3}).
$|n_{10}|$ is obtained from~(\ref{n20_restriction}).
Metric components $n_1$ and $n_2$ are adjusted by an additive constant to match their numerical counterpart:%
\footnote{This is because we have simplified the spike solution by making $F_2(y,z)$ and $F_3(y,z)$ in~\cite{art:Lim2015} as simple as possible. 
The numerical solution again does not necessarily take this simple form. For our purpose here, a zeroth order adjustment (an additive constant) is sufficient.}
\begin{equation}
        n_1 \rightarrow n_1 + n_{1c},\quad
        n_2 \rightarrow n_2 + n_{2c},
\end{equation}
where the value of $n_{1c}$ is approximated by the final value of $n_1$ along $Z=0$, and similarly for $n_{2c}$.
Next, we determine the sign of $n_{10}$, $n_{20}$ and $n_{30}$.
From~(\ref{nonOT_spike_n3}), $n_{10}$ and $n_3$ have the same sign, so we use $n_3$ to determine the sign of $n_{10}$.
From~(\ref{nonOT_spike_n2}), $n_{20}$ and $n_2$ have the same sign, so we use the adjusted $n_2$ above to determine the sign of $n_{20}$. 
$n_{30}$ is obtained again from~(\ref{n20_restriction}). 
$k_2$ is obtained from the relation
\begin{align}
	k_2 = k_1^{-1} \e^{-b^1-b^2} \Big[ &\e^{-2k_0(\tau-\tau_0)} + n_{10}^2 \e^{-\frac12(w-1)^2k_0(\tau-\tau_0)}
\notag\\
\label{k2}
 	& + (n_{10} n_{30} - n_{20})^2 \e^{-\frac12(w+1)^2k_0(\tau-\tau_0)} \Big]^{-1/2}.
\end{align}
$k_3$ is obtained from the relation~(\ref{k_relation}).
This completes the matching procedure for the case $0<w<1$.
Similar matching procedures can be developed for other ranges of $w$.
Note that the matching procedure performs matching along the spike worldline $Z=0$ only,
which is the reason why we need to evolve $\tau$ along $Z=0$ only in (\ref{dtau_dt}).

\section{Results}

As in~\cite{art:Limetal2009}, we shall impose symmetry on the metric components to hold the spike worldline fixed at $Z=0$.
To form a non-moving true spike at $Z=0$, we require that the metric components be even functions of $Z$.

We present a numerical run, in which a generic initial condition similar to that in~\cite[Section IV.B]{art:Limetal2009} is used.
The goal is to show that the joint spike transition observed in numerical simulation is described by the non-OT $G_2$ spike solution.

The format for the initial condition is
\begin{gather}
	\tau=0,\quad	z = \frac12(E_3{}^3)_0 Z + z_\text{zoom},\quad E_3{}^3 = (E_3{}^3)_0,
\\
	\Sigma_1 = (\Sigma_1)_0 \left[ \frac{(3 a_3 a_6 - 3 a_1 a_5 -\sqrt{3} a_5) z^2}{2 (E_3{}^3)_0} + \frac{3(a_4 a_6 - a_2 a_5)z^4}{4 (E_3{}^3)_0} \right]
\\
	\Sm = a_1 + a_2 z^2,\quad	\Nc = a_5 z,
\\
	\Sc = a_3 + a_4 z^2,\quad	\Nm = a_6 z,
\\
	b^1 = (b^1)_0 - \frac{\sqrt{3} a_5 z^2}{2 (E_3{}^3)_0},\quad
	b^2 = (b^2)_0 + \frac{\sqrt{3} a_5 z^2}{2 (E_3{}^3)_0},
\\
	b^3 = (b^3)_0 - \frac{3(a_3 a_6 - a_1 a_5)z^2}{2 (E_3{}^3)_0} - \frac{3(a_4 a_6 - a_2 a_5)z^4}{4 (E_3{}^3)_0},
\\
	n_1 = (n_1)_0 + \frac{a_6}{a_5} \text{exp} \left( (b^2)_0 - (b^1)_0 + \frac{\sqrt{3} a_5 z^2}{(E_3{}^3)_0} \right),
\\
	n_2 =0,\quad n_3=0.
\end{gather}
We use the following values:
\begin{gather}
	a_1 = 1.3,\quad a_2 = 0.002,\quad a_3 = 0.3,
\\
	a_4 = -0.001,\quad a_5 = 0.004,\quad a_6 = - 0.005,
\\
	(E_3{}^3)_0 = 1,\quad	(\Sigma_1)_0 = 10^{-10},\quad z_\text{zoom} = 0,
\\
	(b^1)_0 = 0,\quad	(b^2)_0 = 0,\quad	(b^3)_0 = 0,\quad	(n_1)_0 = 0,
\end{gather}
with $2001$ grid points over the $Z$ interval $[0,10]$ (exploit symmetry and only simulate the right half of the spatial domain), and $T$ interval $[0,40]$.
In particular, $a_1$ is chosen such that we end up with a $w$ value that is not close to $1$, and $(\Sigma_1)_0$ is chosen to be small enough that
we have a distinctive joint spike transition. It takes only a few minutes to run on a single computer.

\begin{figure}
        \begin{center}
                \includegraphics[width=10cm]{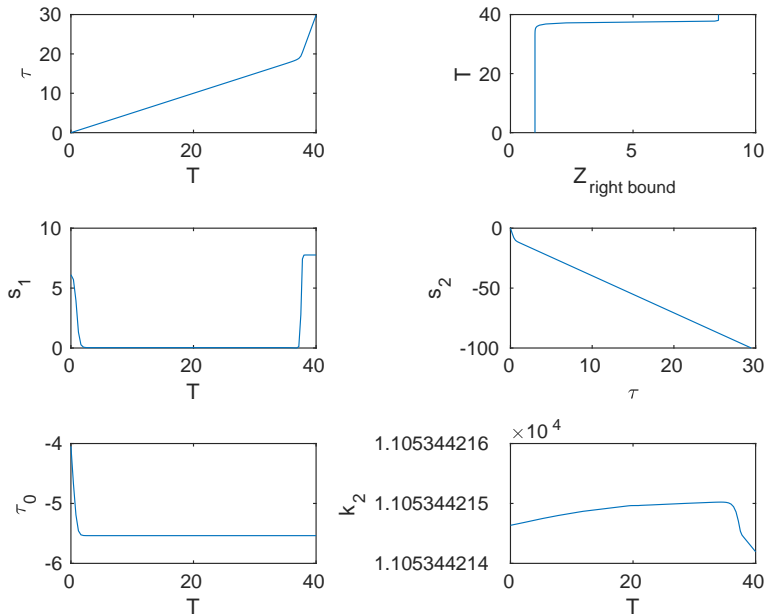}
                \caption{Plots of several variables relevant to the numerical run and matching procedure.}
                \label{various}
        \end{center}
\end{figure}

A joint spike transition is observed after the generic data undergoes a $\Sc$ frame transition.
We use the data from $T \approx 9.8182$ to $T=40$ for matching.
Figure~\ref{various} plots six variables involved with the numerical run and matching procedure, 
namely $\tau$ along the spike worldline $Z=0$ as evolved by Equation~(\ref{dtau_dt}), 
the lower bound for the right-hand excision boundary $Z_\text{right bound}$ given by Equation~(\ref{Z_right_bound}) 
(which shows that the right excision boundary $Z_r=10$ is greater than $Z_\text{right bound}$ for the duration of simulation),
the combinations $s_1(T)$ and $s_2(\tau)$ in Equations~(\ref{sm_sc_combination})--(\ref{temp}), $\tau_0$ and $k_2$ expressions in Equations~(\ref{tau0}) and~(\ref{k2}).

\begin{figure}
        \begin{center}
                \includegraphics[width=\textwidth]{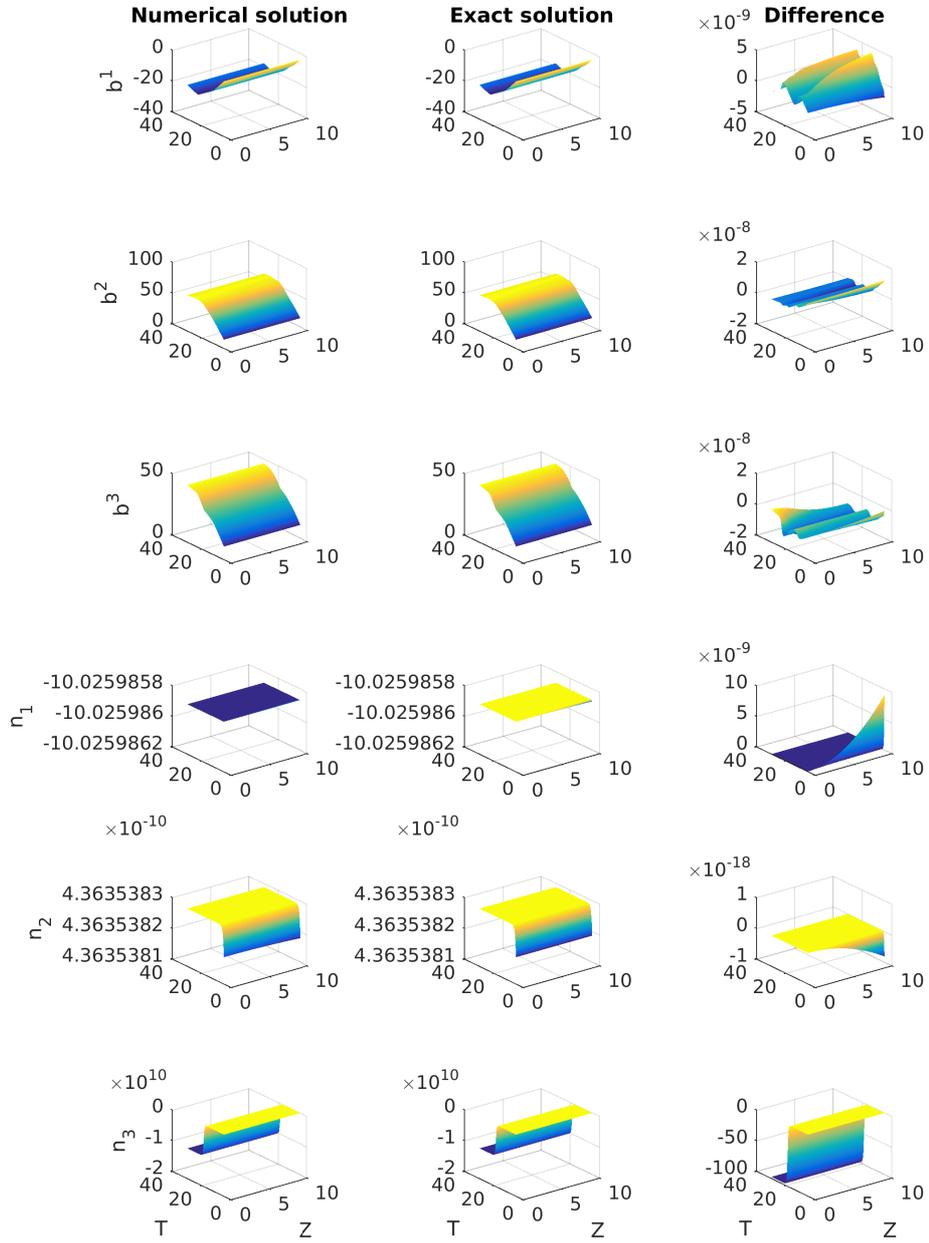}
                \caption{The metric components for the numerical solution, the matching exact spike solution, and their difference.}
                \label{match_generic}
        \end{center}
\end{figure}

\begin{figure}
        \begin{center}
                \includegraphics[width=\textwidth]{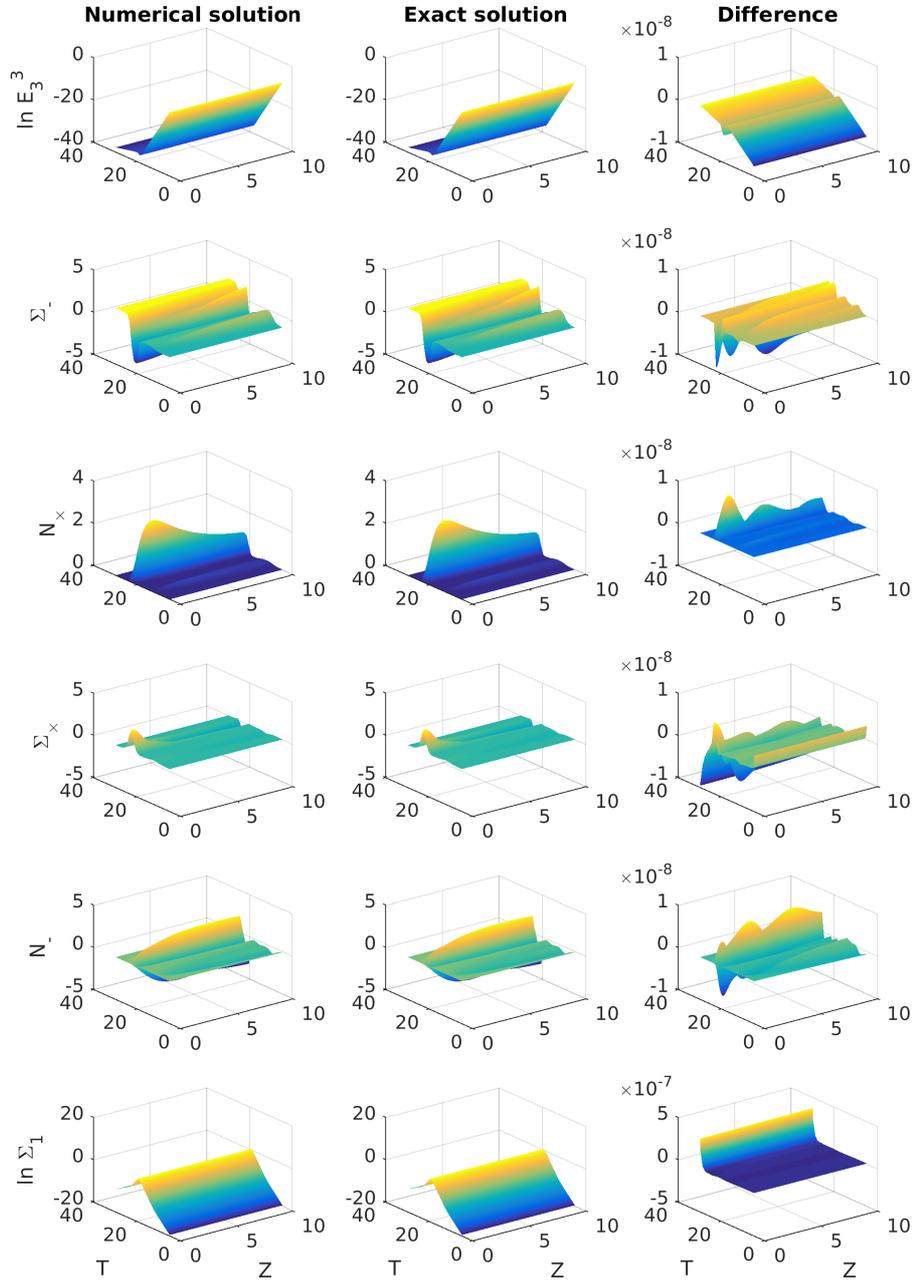}
                \caption{The $\beta$-normalised variables for the numerical solution, the matching exact spike solution, and their difference.}
                \label{match_beta}
        \end{center}
\end{figure}

\begin{figure}
        \begin{center}
                \includegraphics[width=\textwidth]{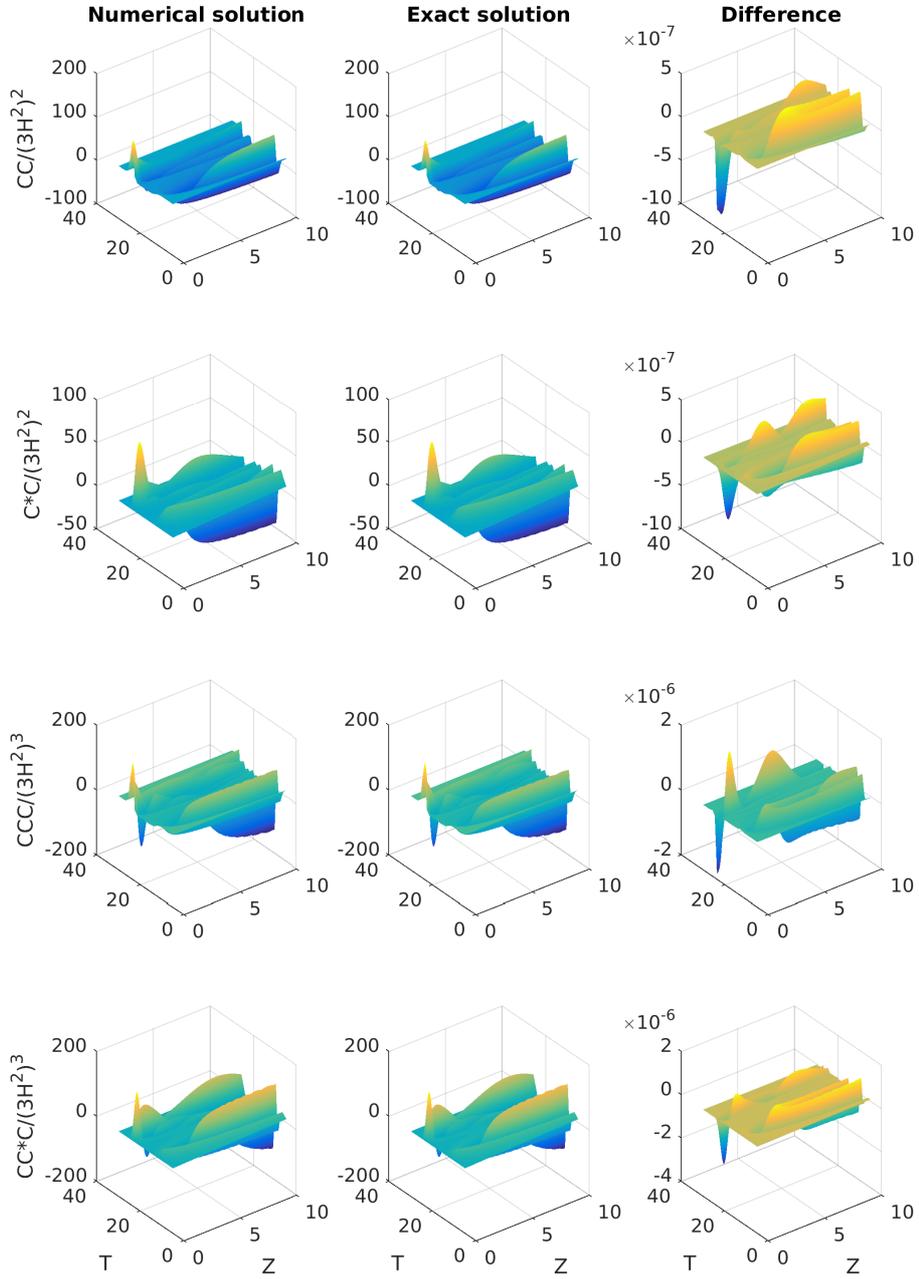}
                \caption{The Hubble-normalised Weyl scalars for the numerical solution, the matching exact spike solution, and their difference.}
                \label{match_weyl}
        \end{center}
\end{figure}

The matching procedure yields the following values (rounded to 4 decimal points):
\begin{gather}
	w = 0.3134,\quad	k_0 = 2.000,\quad	\tau_0 = -5.5372,
\\
	n_{10} = -2.7539 \times 10^{-19},\quad n_{20} = -1.1273 \times 10^{-14},\quad n_{30} = -7.4287 \times 10^4,
\\
	k_1 = 5.8354,\quad	k_2 = 1.1053 \times 10^4,\quad	k_3 = 3.1007 \times 10^{-5}
\\
	n_{1c} = -10.0260,\quad n_{2c} = 4.3635 \times 10^{-10}.
\end{gather}
Figures~\ref{match_generic},~\ref{match_beta},~\ref{match_weyl} plot the metric components, 
$\beta$-normalised variables and Hubble-normalised Weyl scalars (see~\cite[Appendix C]{art:Limetal2009} for their formulas)
for the numerical solution, the matching exact spike solution, and their difference.
They show that the relative difference is at the order of $10^{-9}$ (except for $\ln \Sigma_1$, with difference growing to order $10^{-7}$ at late times), which is quite good.
This is a strong evidence that the joint spike transition observed numerically, even when starting with a generic initial condition, is well-matched by the exact spike solution.
This extends and completes the results of~\cite{art:Limetal2009}. 
While in~\cite{art:Limetal2009} only the Weyl scalars are matched, here we also match the metric components and the $\beta$-normalised variables.

\begin{figure}
        \begin{center}
                \includegraphics[height=8cm]{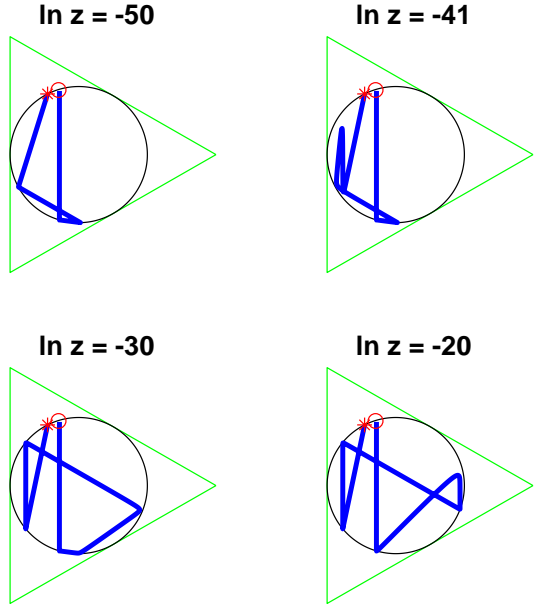}
                \caption{State space orbits projected onto the Hubble-normalised $(\Sp^H,\Sm^H)$ plane, showing a distinctive orbit for each of the four groups of worldlines.
			The representative worldlines used are $\ln z = -50, -41, -30, -20$. A red circle marks the start of the orbit, a red star marks the end.}
                \label{generic_orbits}
        \end{center}
\end{figure}

Recently in~\cite{art:LimMoughal2022}, it was found that the non-OT $G_2$ spike solution has four groups of worldlines with distinctive state space orbits.
To see this in the numerical solution,
four additional numerical runs with $\ln z_\text{zoom} = -50, -41, -30, -20$ and $T$ interval $[0,45]$ are made to produce Figure~\ref{generic_orbits}, 
which shows the distinctive orbits for each of the four groups, projected onto the Hubble-normalised $(\Sp^H,\Sm^H)$ plane. Compare with Figure 6 of~\cite{art:LimMoughal2022}.

\section{Conclusion}

When used in conjuction with the zooming technique for numerical simulations, the area time gauge
avoids the fine-tuning problem encountered in the fluid-comoving or volume gauge used in~\cite{art:ColeyLim2016}.
To recover the time variable of the exact non-OT $G_2$ spike solution along the spike worldline, it is evolved using Equation~(\ref{dtau_dt}).
A new matching procedure is developed for the case $0 < w < 1$. Similar procedures can be developed for the other cases.
We have used a generic initial condition, where the metric components are even functions of $z$ to hold the spike worldline fixed at $z=0$.
Numerical evolution from this initial condition shows a joint spike transition, which is well-matched by the exact spike solution.
This gives a strong evidence that the non-OT $G_2$ spike solution indeed describes the joint spike transition.
This extends and completes the results of~\cite{art:Limetal2009}, strengthening the evidence that spike transitions are part of the generalised BKL dynamics.
The results of this paper can in principle be extended to stiff fluid $G_2$ models.
The numerical code runs efficiently in a typical case ($w$ not too close to 1).
An alternative to zooming technique is the adaptive mesh refinement technique, which requires a high performance computing cluster to run, but which can deal with moving spikes and spikes
beyond $G_2$ models, as demonstrated in~\cite{art:GarfinklePretorius2020}.

\bibliography{}

\end{document}